# Gender Bias in Big Data Analysis

Thomas J. Misa

University of Minnesota

Abstract: This article combines humanistic "data critique" with informed inspection of big data analysis. It measures gender bias when gender-predicting software tools (Gender-API, NamSor, and Genderizer.io) are used in historical big-data research. Gender bias is measured by contrasting personally identified CS authors in the well-regarded DBLP dataset (1950-80) with exactly comparable results from the software tools. Implications for public understanding of gender bias in computing and the nature of the computing profession is outlined. Preliminary assessment of the Semantic Scholar dataset is presented. The conclusion combines humanistic approaches with selective use of big data methods.

Gender bias in computing has immense society-wide ramifications, and our understanding of the troublesome phenomena has expanded dramatically in the past several years. Alongside sharper understanding of power relations in digital devices, networks, and systems, we need deeper understanding of gender bias in computing to correct long-standing societal problems, specifically with women's unusual low participation in computing, at multiple levels of education, workforce, and research, as well as to design better artifacts and systems that do not replicate and entrench outmoded gender norms. Researchers have found that deeply engrained stereotypes about computing can serve to dissuade women from considering a career in computing, and so confronting these societal models and stereotypes is one location for positive intervention.[1]

This article presents a new longitudinal dataset on computer-science researchers (1950–1980) and a critical analysis of big-data analytical tools with the aim of empirically demonstrating and directly measuring gender bias in three popular, highly regarded, and widely used gender-predicting software tools. It starts by reviewing divergent interpretations of gender bias in computing since the 1940s, taking steps to evaluate whether gender bias is a contingent or intrinsic feature of the computing profession. Reform efforts need clear insight whether computing can be altered, and how it might best be reformed; these are not easy matters. It next examines two recent and representative analyses of publications from computer scientists, both deploying immense datasets and each relying on gender "prediction" software tools. The article then combines quantitative and qualitative research methods to evaluate the accuracy of these gender-prediction tools. The results may be surprising to some advocates of computer-driven methods in digital history or digital humanities.  The gender bias in three leading software tools (Gender-API, NamSor, and Genderizer.io) can be analytically measured.

One significant contributor to big-data gender bias is these software tools' use of ahistorical and decontextualized data on ascribed or assigned gender—technically, sex assigned at birth.[2]  Briefly, the tools assume that given names are timeless and unchanging in their gender associations.  The "Leslie problem" is Blevins and Mullen's apt label for gender shifts evident in seven common given names during 1930–2012, with the clear implication that "predicting gender from first names therefore requires a method that takes into account change over time."  They show that Leslie was historically most commonly a male name (in 1900 it was assigned at birth to 92% male babies), then by 1950 its use was split between male (48%) and female (52%) babies, while since around 2000 it is most commonly a female name (96% or greater).  Sydney and Jan were used roughly equally for boys and girls in the 1930s, then Sydney evolved to be an increasingly popular girl's name while Jan initially became a popular girl's name in the 1960s (80% female) then changed to become predominantly a boy's name by the 2000s; a pattern resembling 'Leslie' characterizes Addison, Kendall, Madison, and Morgan which all (at different points in time) evolved from being majority-male to majority-female names.[3]  My preliminary research confirms and extends these findings by identifying 300 names in the SSA dataset that *changed* ascribed or assigned gender across 1925 to 2000, also with a pronounced net "female shift." [see arxiv.org/abs/2210.08983v1 ]

Gender bias in big data analysis using recent data may be subtle and possibly difficult to detect, but in the historical studies examined here we will see that gender bias is substantial.  Other evident weaknesses in some big-data analyses include unwise suppression of "edge cases," where oftentimes significant information is expressed; intermittent misuse of statistical sampling; and periodic inattentiveness to data preparation and data quality.  All such studies rely on inputs of high-quality data; up to 80% of the human labor in big data or data science is proper attention to data identification, aggregation, cleaning, labeling, and augmentation; inattention to any of these dimensions can lead to troubled results.[4]  My conclusion advocates selectively combining the strengths of computer-based quantitative analysis with historically grounded qualitative research to better understand the changing dynamics of gender bias in computing.

It is important to recognize that different scholarly communities and readers of this journal may have profoundly differing understandings of the terms gender, sex, male, and female.  In *Recoding Gender* historian Janet Abbate carefully describes "gender [as] a cultural framework that defines masculinity and femininity as different and unequal," noting that such cultural repertoires are internalized by individuals in shaping their identity.[5]  Gender studies scholars have critically scrutinized these concepts in the past decades, and some may be uncomfortable with this article's effort to count men and women in the computing profession.  While acknowledging the conceptual difficulties in cleanly defining these terms, I believe there is a significant public purpose in gaining accurate knowledge about the number of women and men in the computing profession as well as reasonable confidence about when, and why, the numbers have varied substantially over the past seven decades. As noted above, reform efforts need deeper insight into when and how and why gender bias afflicted computing. I find myself in agreement with Penelope Eckert, who in a 2012 conference paper "Coding for Gender and Sexuality" perceptively assessed the conceptual difficulties raised in computer-centered gender and sexuality studies and yet pragmatically concluded that, "I'd say that [machine learning] corpora should continue with [analysis of] m/f [male/female]" even while recognizing that division as overly simple.[6] Some readers may be unhappy with any "gender binary" since it is an undue simplification of current gender norms and practices. Yet, for historical analysis, I maintain there is merit in attempting accurate counts of men and women in the computing field.

### Emergence of gender bias in computing

Interdisciplinary researchers have examined gender bias in the scientific professions for decades, with scholars such as Margaret Rossiter, Mary Frank Fox, Sally Kohlstedt, Ruth Oldenziel, and many others making significant conceptual and empirical contributions.[7] Gender bias in computing is an atypical case: whereas most scientific, scholarly, and professional fields since the 1980s have become less overtly gender biased in the US and most other wealthy countries—some even modestly female-friendly—computing has instead by numerous measures become more gender biased. Generally, I believe women's participation in computing in the US (by several different measures) grew steadily if slowly from the 1950s through the early 1980s. In the mid-1980s, women earned up to 37% of all US undergraduate computer-science undergraduate degrees and occupied a peak of 38% of the US computing workforce. Rossiter observed that for undergraduate women there was a "collapse ... in computer science ... after 1985."[8] Since that peak year, the proportion of women in computing by most measures fell for two decades or more—prompting wide concern among policy actors like the National Science Foundation and Sloan Foundation, multiple professional society task forces and committees, and not least the computing profession itself.

Researchers flooded in to help understand these worrisome developments. The immense volume of social science research was assessed by Joanne McGrath Cohoon and William Aspray in their 2006 edited volume, *Women and Information Technology: Research on Underrepresentation*, which somberly concluded at that time "twenty-five years of interventions have not worked."[9] A low point was reached in 2009 when the highly regarded Computing Research Association–Taulbee annual survey found women gaining just 11.2% of undergraduate computer science degrees. Ever since the mid-1980s many smart people have devoted significant time, effort, and resources to correct gender bias in computing, and there are some success stories such as the annual Grace Hopper Celebration of Women in Computing.[10] But for the computing profession at large there are no easy answers in sight.

Recent critics of "algorithmic bias" in computing have shown one of that phenomenon's effects to be gender bias. Scholars such as Cathy O'Neil, Meredith Broussard, Ruha Benjamin, Safiya Umoja Noble and others have identified two general modes of troubling bias: one, when there is a "real-world" object that an algorithm purports to measure or predict, and gets it wrong; the other, when consequential determinations such as school admissions, prison sentencing, job application screening, or consumer credit reports are made by algorithm absent any alternative "real-world" results that might be used for evaluation.[11] (This present article examines data where a "real-world" and algorithmic result can be directly compared.) A recent *Information & Culture* review praises Caroline Criado Perez's *Invisible Women: Data Bias in a World Designed for Men* (2019) for documenting that "gender data gaps are pervasive, relentless, and dangerous" as well as drawing attention "to empirical issues of gender and data bias."[12] The argument of Mar Hicks' *Programmed Inequality* (2017), winner of the American Historical Association's 2019 Herbert Baxter Adams Prize in European History, is reprised in their essay "Sexism is a Feature, Not a Bug" in the purposely polemical *Your Computer is On Fire* (2021), where chapters by Corinna Schlombs, Kavita Philip, and Safiya Umoja Noble also critically examine gender issues.[13]

There is wide agreement that gender bias in computing is bad, but there are divergent views about *when* it emerged—and, consequently, *whether* and to what extent gender bias is intrinsically and permanently embedded in the computing profession. The statistics cited above showing women's growth in computing during roughly 1965–1985 are not universally acknowledged. Indeed, a view that gender bias is inherent and intrinsic to the computing profession has become the dominant accepted view; this interpretation is widely cited in diverse scholarship and has been publicized in varied popular media including the Smithsonian, National Public Radio, *Wall Street Journal*, *Atlantic Monthly*, *Fast Company*, and numerous blog posts, as well as Robin Hauser Reynold's acclaimed 2015 documentary "Code: Debugging the Gender Gap."[14] This perspective was anchored in 1999 when historian Jennifer Light observed that beginning with the pioneering women who programmed the digital computer ENIAC, unveiled in 1946, women took up notably prominent roles in early computing. Light pointed to an idiom of sex-typing: "designing [computer] hardware was a man's job; programming was a woman's job"—and went on to suggest "how the job of programmer, perceived in recent years as masculine work, originated as feminized clerical labor." Nathan Ensmenger, expanding this insight, suggested in two prominent essays that in the 1950s women comprised likely 30% of early computer programmers (and possibly as much as 50%) but that in subsequent decades programming was "made masculine" when men desired to remake computing as a high-status professional field. His assertion that "the masculinization of computer programming" during the 1960s and early 1970s produced a pervasive and permanent masculine culture in computing, ever since, received wide recognition, not least through Reynold's documentary film and numerous popular media accounts.[15] The "making programming masculine" perspective has been supported by compelling anecdotes, advertisements embodying *Mad Men* style sexism, and theoretical analysis of (male-driven) professionalization, advancing the argument that computer programming was initially a female-dominated field with men subsequently pushing women out during professionalization in the 1960s and 1970s.

In contrast, data-rich or longitudinal studies have most often identified *growth* in women's participation in computing in the 1960s and 1970s rather than male-driven exclusion of women. As noted above, NSF data on computer-science undergraduate degrees and Census Bureau workforce data indicate women's growing participation in computing through the mid-1980s. One 2011 study found that women researchers authored an increasing proportion of computer-science conference papers, expanding from 7% of authors in 1967 to 27% in 2009. A 2015 article found that women were an expanding proportion of published computer-science authors, increasing from around 3% of authors in the mid-1960s to 16% by 2010. My research in professional membership rosters and computer user groups also indicates growth in women's participation in computing from the 1960s through 1980s.[16] Abbate's view, based on 52 interviews and her wider analysis of the field, is that "women . . . held positions of responsibility and influence in the early computer industry . . . and were employed in numbers that, while a small minority of the total, compared favorably with women's representation" in other areas of science and engineering at the time. Agreeing that "openness of programming jobs to women during the 1950s is often exaggerated," Haigh and Ceruzzi in their recent survey of the field state "there is no evidence to support the widely repeated assertion that 30 to 50 percent of all programmers were women during the 1950s and 1960s, or that this proportion dropped during the 1970s."[17] The view centered on these large-scale studies indicates growth—not decline—in women's participation in computing research, publications, education, workforce, and professional activities from the 1950s through the mid-1980s. This second view suggests that gender bias is not an intrinsic, inherent, or permanent feature of the computing profession but instead it was contingent and so is reversible. Much hangs on which of these two perspectives on gender bias is accepted.

I know of only two datasets that might support or suggest a *decline* of women's participation in computing during the 1950s or 1960s. One is reported in *Solving the Equation: The Variables for Women's Success in Engineering and Computing*, a widely circulated 2015 study from the American Association of University Women (AAUW), which cites US Census data on the workforce and appears to show a women's decline in "computer and mathematical occupations" from 27% in 1960 to 20% in 1970 before growth resumed through 1990. Yet a note cautions the use of Census categories: "For computer and mathematical occupations in the 1960 census, no category for computer scientists was included; in the 1970 census, the category was titled 'mathematicians and computer specialists'; and in the 1980, 1990, and 2000 censuses, the category was titled 'mathematical and computer scientists'."[18] In the 1960s insurance companies employed a large number of mathematics majors, and many women, as actuarial staff, expanding with the growth of life insurance and credit insurance.[19] Again, not until the 1970 Census is there solid information on women in the *computing* workforce.

The second study that might suggest a decline of women in early computing comes from four researchers at the Allen Institute for Artificial Intelligence who examined 11.8 million computer-science publications beginning in 1950. Their data and their methods are analyzed in the next section below. Advocates of the 'making programming masculine' perspective may be heartened by their findings about computer science research, especially during 1950–1970. Based on the DBLP computer-science research dataset, and supplemented by the even-larger Semantic Scholar dataset that the Allen Institute created in 2015–17—and relying on gender-prediction software—the authors form a time-series graph plotting the proportion of women authors in computer science publications across time. In their graph, women computer-science authors start at 20% in 1950 (the first year their study tabulates DBLP's coverage) and climb briefly to about 22% the following year; then the figure for women authors drops, slowly if unevenly, to 15% in the early 1960s and slides further to around 13% in the late 1960s.[20] Their large-scale data, if confirmed, would be consistent with and empirical support for two key claims frequently made by advocates of 'making programming masculine': namely, that women were a numerically large proportion of the early computing community (20% in the early 1950s is the highest such *systematic* data point known to me); and that across the 1950s or 1960s it transpired that women left the computing field, possibly as men re-shaped the field during its years of professionalization.

**Big data in historical gender analysis**

The current fashion for big data and the enthusiasm for digital humanities has resulted in large datasets being used to define and diagnose gender bias in computing. Our knowledge about gender bias is in part shaped by big data analysis. Traditional narrative history is perfectly appropriate when examining small populations, and when populations become larger than two dozen or so historians can use collective biography, or prosopography, for up to around 100 persons.[21] Beyond this, however, traditional historical research methods—involving fine-grained research in primary-source archival materials, reconstruction of individual experiences, based on highly disaggregated historical data—are often unwieldy or unworkable. A recent article in *Information & Culture* by Nooney, Driscoll, and Allen uses mixed methods to assess 1,285 published letters to the editor of *Softalk* magazine during its "short reign" (1980-84) to gain insight into ordinary or non-elite *users* of personal computers. The authors deftly combine traditional historical research with "software-assisted methods" to assess the gender, location, and topics of concern for the population of letter writers, including cautious use of "the automated judgment of a gender classifer program." "We did not use automated gender inference as a replacement for but rather as a complement to human judgment," they note.[22]

Published gender research with datasets larger than several thousand typically rely extensively or even wholly on software tools for determining authors' gender. Three of the most widely used such tools are Gender-API, NamSor API, and Genderizer.io. Their use by diverse researchers has led to a substantial cottage industry of computer-savvy assessment researchers to evaluate them. "Gender API is in general the best performer in our benchmarks, followed by NamSor," according to one assessment.[23] Blevins and Mullen critically scrutinize Genderizer.io, finding it "unsuitable for historical work . . . because it is based only on contemporary [given name] data," with similar liabilities in the widely used Natural Language Toolkit for Python programming.[24] The following paragraphs examine the research methods used in two recent computer-science gender studies that used NamSor and Gender-API. This article's direct assessment of the three software tools follows.[25] Multiple published studies have used these tools as a principal means to assess gender in varied populations.[26]

Women's persistent under-representation in computing has prompted scrutiny of the "gender gap" in computer-science research. The ambition of eventually calculating "the overall proportion of women in computer science as a whole" informed one focused study by five German researchers who deployed a bibliometric approach and used NamSor API to gender identify names selected from the DBLP computer science database. DBLP aims to cover computer science research publications, beginning in 1936 through the very near present. With its origin in France, NamSor uses a database of 5.5 million names to predict gender, assess cultural origins, and map the economic flows of globalization while offering worldwide coverage including "all languages, alphabets, countries, regions" including diverse language scripts. The German researchers are clear enthusiasts. They write that NamSor's "innovative machine learning algorithm provides unmatched accuracy at a fine-grained level," and they used it to assess gender in a refined sample of 13,101 names from 2012 to 2017. They, too, like Nooney et al., manually verified a select sample of around 300 female and 370 male names, using internet searching to gender-identify the individual sampled authors, as a means to evaluate the automated results.[27] The results were mixed. Whereas NamSor and their manual verification agreed on 84% of the male names (that is, actually being male persons), with just 0.3% being wrongly identified as female; the female names were correctly identified in only 70% of the cases, with 17.6% being misidentified as male. The sad news, regardless of the fine-grained analysis, is that women were just 10% of researchers in this specific sub-field of computer science.[28]

An immense big data study by the Allen Institute for Artificial Intelligence (noted above) also examined the DBLP database as part of the larger Semantic Scholar dataset that the Allen Institute itself created.  Examining 11.8 million computer-science publications, the authors acknowledge dependence on "Gender API to perform gender lookup for each name."[29]  Gender API was created by a German social-media programmer in 2014, originally to automate and streamline online registration forms, and its current database has 6 million names from 189 countries. These authors praise Gender API's "large online database of name–gender relationships derived by linking publicly available governmental data with social media profiles in various countries."  Similar to NamSor, for each first name "Gender API outputs the predicted binary gender (*female* or *male*), along with the accuracy associated with the prediction and the number of samples used to arrive at that determination." (Actually, Gender API has a flawed conception of "accuracy."[30])  These authors recognize that many names are not precisely gender binary, so they pragmatically adopt a "simplified view of gender as a probability distribution over two genders" (male and female).[31] No sampling or manual verification or other validation of the automated gender-prediction results is reported; the results rely wholly on automated gender prediction by Gender API.  This study, as noted above, suggests the tantalizing finding for the early period in computing (1950s) of around 20% women as research authors.[32]

With their big data machinery in place, the Allen Institute researchers extended their gender analysis to an additional 140 million bibliographic entries (1940–2020) ranging across 19 major scholarly disciplines defined by Microsoft Academic.  These include science fields like biology, chemistry, environmental science, geology, materials science, mathematics, and physics; social science and humanities fields including art, economics, geography, history, philosophy, political science, psychology, and sociology; and professional fields like business, medicine, and engineering; and of course computer science.  The results are plainly incredible.  Perhaps, as Brian Beaton remarked about the challenge of data science, such studies "have the ability to capture and reflect our worlds in deep and captivating ways that were not previously imaginable."[33]

In a later section below, I will take up Beaton's suggestion following cultural critic Lionel Trilling, to engage in "data critique" from an "emphatically humanistic practice thoroughly rooted in arts and letters" but first, in a complementary spirit, I will scrutinize the analytical and mathematic methods used in this big data analysis.  My research goal is qualitative understanding of women in computing; my methods here are informed inspection of big-data analysis.  I aim to inquire into the "semblance of neutrality and objectivity" that typically accompanies such immense data-rich studies.[34]  In this case, several of the historical findings of the Allen Institute's study are likely to be unsound.

Here are two striking findings, each in their way incredible.  In the field of medicine, the Allen Institute found the proportion of women researchers begins in the 1940s at 15%, growing by 1947 to greater than 20%, a level sustained for several decades, finally climbing to a peak of just over 30% by 2018. For the field of engineering, the Allen Institute found women to be 10% of research authors in the 1940s, climbing during the war years to 15%, then sustaining a range of 10-13% from the 1950s through the 1970s, again rising to a recent peak near 30%. The other 17 scholarly fields have at minimum 10% women researchers; many have substantially more. Have historians of women in the technical professions missed something?

Or did this big-data analysis somehow send us astray? In the US, we know that in medicine as a profession only 6 percent of the workforce were women in 1950, with men thoroughly dominating all medical specialties and nearly all research positions; medical schools started admitting significant numbers of women only in the 1970s and not until around 1980 did *new* medical graduates finally reach 20% women.[35] Like medicine, engineering was soundly male-dominated during these decades; so that "as late as the 1960s, women still made up less than 1 percent of students studying engineering in the United States."[36] It is literally unbelievable that published engineering researchers might be 10% women, published medical researchers 15% women, or published computer science researchers 20% women *anytime* in the 1950s. To better understand some possible roots of these anomalous findings, the following section presents my detailed analysis of the DBLP dataset contrasted with the results from the three gender prediction software tools. Preliminary discussion of the surprising divergences of the Semantic Scholar and DBLP datasets follows.

**Measurement of gender bias**

To assess these big-data findings and methods, and to directly measure gender bias in the use of gender-predicting software tools for historical research, I formed a sample dataset of DBLP entries from 1950 to 1980 (data extracted May 2021). For the first four years (1950-53), I can identify 87 percent or more of the DBLP authors individually; for 1960, 1970, and 1980, more than 80 percent. For the remaining roughly 10-20% of authors I computed a year-specific gender probability lookup using the Social Security Administration's comprehensive database of all given US names since 1880.[37] I have used a similar method in two preceding articles, and will briefly review the refinements that were helpful in analyzing slightly over 10,000 computer-science research publications.[38] For several years, I have been using multiple and varied data sources—including lists of conference attendees, professional membership rosters, and computer user-group conference attendees—to estimate the proportion of women in computing from the 1950s through the 1990s; these include the decades prior to the availability of the 1970 US Census figures for women's participation in the US computing workforce. My goal has been to compute women's participation rates, as a percentage of these various communities. Others have used the DBLP dataset to investigate classic bibliometric questions about research productivity, co-authorship, and co-citation analysis.

The early years of the DBLP dataset are a glimpse into the "pre-history" of computer science.[39] In the 1950s, the research articles in DBLP are drawn from symbolic logic, operations research, with some electrical engineering, statistics, telephone engineering, and glimmers of information theory. In 1950, no less than 26 of the 28 total entries are in symbolic logic with the vast majority from the singular *Journal of Symbolic Logic* (founded 1936) with three articles from the *Archive for Mathematical Logic* (founded 1950). Two books rounding out that year are *The Human Use of Human Beings: Cybernetics and Society* by Norbert Weiner and *Statistical Decision Functions* by Abraham Wald. All 28 authors, many truly distinguished figures, can be clearly and personally identified. Two of this number were women: Ruth Barcan Marcus and Rózsa Péter, and each of them have Wikipedia entries, as do most of the other 1950 authors.

The years 1951-53 continued reliance on *Journal of Symbolic Logic*, while drawing additionally on operations research and other emerging computing areas. Tables 1 and 2 show that more than 80% percent of DBLP authors during 1950–1980 can be *personally* identified; these are not 'names' but individual persons, well known in the field and recoverable today (see note on method below). For the remaining 20% or fewer authors, I supplemented personal identifications with name- and year-specific queries of the SSA database, selecting as 'year of birth' a point 30 years prior to the publication's year; thus for articles in 1953, I looked up authors' first names in the 1923 SSA data. For each name, year by year, the SSA tabulates the total number of instances it was used for a male child and the instances for a female child. For many names, the probability of the name's being a female name—let's say p(F)—clustered around 0 to 0.05 for many male names (such as Carl, David, John, Robert) or around 0.95 to 1.0 for many female names (such as Anna, Harriet, Mary, Nancy, Patricia). But a substantial number of names, especially in the later decades, simply do not fit into the "gender binary," including Arie 0.61, Augustine 0.29, Willy 0.18, Chris 0.09, and Ian 0.08, citing their respective p(F) values. These latter names—indeed all names—can be systematically tabulated with their associated p(F) values to form an inclusive dataset.

Some computer-science gender studies, assuming gender to be binary, unwisely remove any 'name' with a gender probability more than 5% outside the binary norm — resulting in the unsound disregard of all the discarded individuals. This flawed practice suggests one way that gender bias might be resident in big data studies: suppression of edge cases. In 2016 Kamil Wais targeted "The Role of Gender in Scholarly Authorship," a data-heavy PLOS study where the authors, remarkably enough, "decided *a priori* to use only those records that have at least a 95% probability to correctly predict gender based on a first name." Affirming that very point, the authors provided this excuse: "Otherwise, as with 'Leslie' or 'Sidney', we are unable to identify the gender and do not include that author in our analysis." This discarding has the worrisome consequence of banishing such well-known and prolific authors in computer science as Leslie Valiant and Leslie Lamport (the male Turing laureates in 2010 and 2013, respectively) as well as the female computer scientists Leslie Ann Goldberg and Leslie Pack Kaelbling (full professors at Oxford and MIT, respectively). My preliminary research using the ACM Digital Library identifies no fewer than 133 computer scientists named Leslie active from 1970 to 2000 [see [arxiv.org/abs/2210.08983v1](arxiv.org/abs/2210.08983v1) ] . Of them 37 were men and 83 were women (the gender of 13 could not be identified)[40]— and all of them were excluded from the above gender-binary study.[41] Digital humanities studies using natural language processing, machine learning, and other computer-driven modeling and analysis techniques also sometimes suppress edge cases and very likely weaken the robustness of their findings.[42]

    A reader may wonder if reliable information can be obtained about these thousands of authors located in DBLP. With the rise of social history in the 1970s, historians developed research tools such as linking archival records and deep biographical lookups of non-elite persons. Traditionally, any medium-sized population can be analyzed (e.g. for 'collective biography' as noted above) using well established reference tools. These include locating biographical data from newspapers, magazines, photographs, public events, career summaries, correspondence, testimony of family or colleagues, obituaries, and sometimes a bit of luck. Ample personal data also can be found today using Google searches, Linked-In's career summaries (and gender-revealing recommendations), social media sites, personal and institutional websites, and the Internet Archive. Some computer science figures can be tracked through the internet community's websites. Thousands of author bios appear in IEEE publications.

    My basic *modus operandi* was to find an individual's institutional affiliation or geographic location using the ACM or IEEE digital libraries (alas DBLP does not readily list such metadata) then opportunistically conduct searches in WorldCat, Google, published author biographies and encyclopedias, and multiple other sources. While there is documented concern about gender bias in Wikipedia, I found many accomplished women computing researchers there and at least one with a passing reference to her colleague-husband (with no entry of his own).[43] Even "initials-only" names can frequently be personally identified. Among others, I found R. M. Martin in Wikipedia to be logician Richard Milton Martin; J. H. Woodger to be philosopher Joseph Henry Woodger; and J. M. Bennett to be computer scientist John Makepeace Bennett. I emphasize Wikipedia was only one source to personally identify the DBLP authors. Additional "initials only" authors were identified through the ACM and IEEE digital libraries, WorldCat, institutional and personal websites, authors' bios, and Google searches. The ACM DL provides a means to track individual authors across their careers, identifying persons who may have published using slightly varied names, or initials-only names, or who change last names.[44]

Through these varied sources, I was usually able to locate multiple independent sources of gender-identifying information; individuals for whom I did not locate positive personal information remained in the unknown-gender category, and were included in that year's population total but not in the tabulations of (probabilistically) male or female authors. Some authors of course might purposely choose to keep information on their gender out of their public profiles. Another general case where my method was quite successful, through 1980, was with East Asian-named authors, who have been removed from some gender research datasets owing to the difficulty their names cause for gender-predicting software—another instance of suppressing "edge cases" in a dataset.[45]

For each year I created a population of the gender-identified persons publishing that year — rather than a tabulation of the total research publications. This permitted me to more fairly estimate the total number of women in the population of computer scientists, and assess how this figure changed across time; it also allows comparison with earlier research on computer-science populations. To make this clear, tables 1 and 2 indicate percentages of women researchers (in the *population* of researchers that year). Apart from the singular year of 1950 when two women were among the total population of 22 authors, women's participation in early computer science (at least as expressed in the DBLP dataset) was dreadfully low. In 1951 and 1952, there were no women authors; while in 1953 the sole woman author was (again) Ruth Barcan Marcus, whose "Strict Implication, Deducibility and the Deduction Theorem" appeared in *Journal of Symbolic Logic*. Across these four years, then, there were three women annually appearing among the DBLP's 173 identified authors. How this figure of about 1.7%—relying on positively identified individual persons rather than software analysis of 'names'—becomes the Semantic Scholar's 20% women authors in computer science is somewhat mysterious. Recall that the Allen Institute used Gender API as its sole means to identify gender in its analyzed authors, and so the software tool's gender bias (noted below in Table 4) could contribute to these anomalous results.[46]

**Table 1: Women DBLP authors (1950–1953)**

| Year | Total (articles) | Women authors % | Total population (authors) | ID = person % |
|---|---|---|---|---|
| 1950 | 28 | 9.099% | 22 | 100.0% |
| 1951 | 25 | 0.064% | 24 | 91.7% |
| 1952 | 40 | 0.091% | 33 | 90.9% |
| 1953 | 128 | 1.158% | 94 | 89.4% |

**Table 2: Women DBLP authors (1960–1980)**

| Year | Total (articles) | Women authors % | Sampled population (authors) | ID = person % |
| --- | --- | --- | --- | --- |
| 1960 | 484 | 2.702% | 235 | 85.5% |
| 1970 | 2,039 | 3.108% | 341 | 83.6% |
| 1980 | 7,359 | 4.078% | 697 | 82.4% |

For larger populations such as the thousands of authors in 1970 and 1980, different research tools are needed. Let's imagine you want to know the number of right- and left-handed and ambidextrous persons attending a soccer match in Madrid's immense stadium with seating for 100,000 fans. You don't need to interview each of them. Nor can you get a fair estimate from counting people in one of the locker rooms; but you might form a statistically valid sample from across the stadium, and assess this much smaller number of persons. Pollsters, census officials, social-science researchers and many others deploy statistical sampling all the time. From well-known formulas, taught in undergraduate statistics classes (but possibly unfamiliar in computer science[47]), you can make a reasonable estimate of large populations — with allowance for (say) a 95% confidence level of being within 5% of the true figure — if you form a sample of 383 persons chosen at random. For the more modest numbers of DBLP authors one can reduce the statistically valid sample size: for the original 620 total named authors in 1960, a valid sample size is 238.

Statistical sampling offers a powerful and robust alternative to the big-data approach of pouring every case through an algorithm (and, possibly, the temptation of trimming troublesome edge cases). I can testify that samples of 300 or so individual names are feasible to identify personally; I formed a sample of each year's statistically chosen articles, then analyzed these articles' authors, and co-authors if any, as separate individuals. The number of *articles* and the number of *authors* is not the same of course, owing to some authors publishing more than one article in a year as well as some articles having multiple authors.[48] Figure 1 presents an overview of the 1950–1980 gender-identified DBLP data set, with the x-axis being years, the y-axis being tabulated women's participation (as *authors*), and the bubble area sized as the number of *articles* published that year. (Owing to the small populations and diverse results during 1950–53, with, respectively, 2, 0, 0, and 1 women), I combined these four years into one composite data point representing 221 publications and 173 individual authors. A computed linear trendline clearly indicates growth across these decades, and a respectable $R^2$ figure of 0.97 indicates good "fit" of the trendline to the underlying data points. Figure 1's overall picture of women's growing authorship in computing research across 1950–1980 generally agrees with my earlier published data based on ACM conferences, membership rosters, and computer user-group attendees.  Even if DBLP might include an unduly small number of women in its population of computer-science authors (as discussed below), women's participation as computer-science authors clearly increased during these decades.

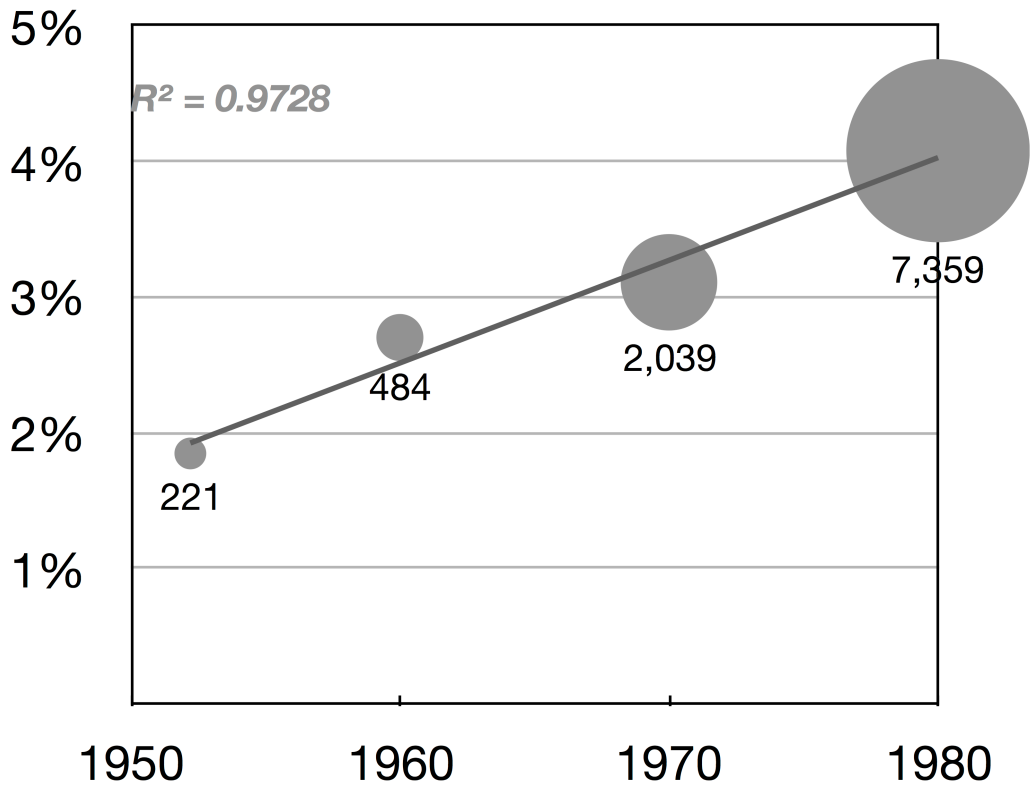

Figure 1: Women's participation in DBLP dataset

There are distinct forms of "gender bias" that may be found in big-data studies.  The critical literature on algorithmic bias directs attention to instances where women (and other non-dominant groups) are not recognized, or under-counted, or marginalized in some way.  Let's call this "type one" gender bias (I will discuss type two bias shortly).  With my dataset, it can be exactly located by contrasting the actual identified persons in DBLP (noted above) with the algorithmic results if a researcher were to use one of the gender-predicting software tools discussed above.  I have done both methods, and the results are quite something (see table 3).  Here are instances of type one bias resulting in female under-counting: In 1960 Mandalay Grems published two articles in *Communications of the ACM,* including a report on the bellwether IBM user-group SHARE; 1970 included Bell Laboratories mathematician Florence Jessie MacWilliams, later co-author of *Theory of Error-Correcting Codes* (1983); Purdue University professor of mathematics Jean Estelle Rubin, author of five books on axiom of choice, set theory, and mathematical logic; Harriet H. Kagiwada, an expert on women in the professions cited in a National Academy report; and mathematician–systems engineer Love H. Seawright, cited for her contributions to virtual machine development at IBM Cambridge Scientific Center; while 1980 included Shigeko Seki, a computer-science PhD from Kyoto University who specialized in cellular automata, formal languages, and algorithms while a professor at Cal State–Fresno; Joan Francioni, a specialist in parallel computing who received the first computer science PhD awarded by the University of Florida and later worked with visually impaired students; and Mildred S. Joseph, an ACM author and computer specialist at the University of Texas.

**Table 3: 'Type One' Gender Bias in Gender-Predicting Software Tools (1950-1980)**

| Name | ID as *female* | p(F) GAPI | p(F) NamSor | p(F) Gen-IO |
| --- | --- | --- | --- | --- |
| Mandalay Grems | Local obituary | **0.42** | **0.51** | **0.58** |
| Florence Jessie MacWilliams | NYT obituary | 0.84 | 0.86 | 0.79 |
| Jean Estelle Rubin | Wikipedia | **0.11** | **0.12** | **0.05** |
| Love H. Seawright | IEEE DL | 1.0 | 0.83 | **0.59** |
| Joan Marie Francioni | Google-info | **0.44** | **0.52** | **0.31** |
| Shigeko Seki | Fresno State | **0.02** | **0.59** | 1.00 |
| Harriet H. Kagiwada | NAP report | 0.96 | **0.69** | 0.97 |
| Mildred S. Joseph | Univ Texas | 0.97 | **0.64** | 0.97 |

Table 3 shows that each of these accomplished and personally identified women would be substantially misidentified using gender-predicting software tools.[49]  In my tabulations I set p(F) at 0.9999 for real-world, positively identified women; many times, the software tools return reasonable figures (p > 0.9) for women's names.  But the software tools return p(F) figures for these women's names as low as 0.02, clearly in error but otherwise and elsewhere an unquestionably "male" figure.  For table 3's positively identified women, the median figure for p(F) = 0.615.  Additional forms of type-one gender bias occur when women researcher-colleagues were relegated to 'acknowledgements' while their male colleagues claimed authorship, or other instances of hidden or invisible authorship; when there is unwise suppression of edge cases, such as names not conforming to gender norms resulting in the disregarding (e.g.) of female computer-science authors named 'Leslie'; and with the gender bias resident in Wikipedia entries. For the DBLP dataset itself, there may be additional gender bias that results from its selecting which articles in the professional published literature are included in DBLP. I am presently investigating its selection of articles from the annual conferences of several of ACM's Special Interest Groups; the one devoted to University and College Computing Services, or SIGUCCS, featured a great many female authors. Further analysis is needed to establish to what extent such selection might contribute to significant gender bias in the DBLP dataset itself and how this bias might change across time.

The 'type two' form of gender bias in big data will not be so familiar. Type two does not suppress or ignore women, but it does distort the historical record and lead to incomplete or erroneous understanding of men's and women's changing participation in computing. As noted above, an unrealistic conception of women's numerical predominance in early computing—reportedly 30% to 50% women programmers—has contributed to a widespread scholarly and popular view of computing and computer science as intrinsically and permanently biased against women. 'Type two' bias can result when research methods under-count historical male persons, owing to the substantial "net positive" female-shift in US names across the decades in study, as noted by Blevins and Mullen. As mentioned above, most gender-prediction tools are focused on—one might say trained on—today's existing name–gender associations. Historically, though, many names had different gender associations, and the preponderant shift in the mid-twentieth century was for many historically male names to be transformed partly or wholly into female names, such as the computer science Leslies.[50] While table 3 identifies eight women likely to be misidentified by the software tools as men, the countervailing type-two gender bias is expressed with scores of male authors who are partly or wholly misidentified as women. Here are ten instances, the first four logicians: Hao Wang has gender-predicted p(F) values ranging up to 0.23; Schiller (Joe) Scroggs, 0.28; Jan Kalicki, 0.31; Shen Yuting, 0.35; analog computer specialist Lee Cahn, 0.49; linguist Noam Chomsky, 0.31; information theorist Chacko Abraham, 0.29; physicist Satosi Watanabe, 0.39; mathematician Dominique Perrin, 0.55; and, not least, aerospace-industry executive Carmen Palermo, up to a securely female-sounding 0.98. There are dozens and dozens more. It is time to shift the level of analysis from individuals to the population in the dataset as a whole.

An overview of systematic gender bias in the three software tools is presented in table 4. It tabulates results beginning in 1950 and extending to 1980. For each year, I have normalized the results by dividing each software-tool's computed result by the corresponding personally identified "real world" result (see table 1). I limited this analysis to the "matched" individuals with both personally identified as well as software-computed p(F) figures; the ratio is an exact one comparing the *same individuals* identified by *different means*.[51] If there were perfect agreement, in each year, the figure shown in the three columns for Gender API, NamSor, and Genderizer.io would be 1.00. The table demonstrates that the software gender tools consistently predict figures for women's participation that are surprisingly and erroneously high. Across all results in the three columns, the *median* of the diverging ratios is 2.13 suggesting the gender tools' results are something like twice the proper "real world" value (while the *average* of the diverging ratios is skewed up to 14.6 owing to the several massively inflated figures, ranging from 25-fold up to 70-fold across the three software tools).

**Table 4: 'Type Two' Gender Bias in Gender-Predicting Software Tools (1950-1980)**

| Year | Personal ID | GAPI ratio | NamSor ratio | GenderIO ratio | N(match) |
|------|-------------|------------|--------------|----------------|----------|
| 1950 | 1.00 | 1.25 | 1.42 | 1.19 | 21 |
| 1951 | 1.00 | 39.73 | 59.21 | 37.74 | 22 |
| 1952 | 1.00 | 36.01 | 71.64 | 31.71 | 28 |
| 1953 | 1.00 | 2.73 | 3.52 | 3.14 | 80 |
| 1960 | 1.00 | 1.72 | 2.07 | 1.78 | 182 |
| 1970 | 1.00 | 1.78 | 2.39 | 1.85 | 325 |
| 1980 | 1.00 | 1.69 | 2.13 | 1.71 | 668 |

The purpose of this analysis was to analytically measure gender bias in these three software tools. Perhaps I can offer some suggestions as to why these gender tools are so gender-biased: off from real-world values by as much as 70-fold. It would be a serious matter if any software tool might predict an average North American human's weight inflated to 12,000 pounds! As noted above, many historically male names—here actual computer-science authors—are systematically misidentified as female names by the gender-prediction tools' use of today's name–gender associations. There is of course both 'type one' and 'type two' gender bias; but type two predominates in this dataset. There is no worrisome suppression of 'edge cases' since I analyzed all individual computer-science authors in four ways (identifying them personally, with 80-100 percent success, and then doing gender-prediction with each of the three software tools). The most extreme results come for the two years when the actual real-world data has zero women, and yet the gender tools return non-zero probabilities for many of the all-male names; these add up.  For instance in 1951, the three tools' calculated p(F) figures are 0.0267, 0.0408, 0.0260 (not terrible for all men) but my individually identified p(F) figure is much smaller at 0.0006.  (Recall the latter figure is the denominator in computing the normalized gender-bias ratios; all computations done with spreadsheet accuracy and not these truncated figures.)  Similarly in 1952, the respective p(F) figures are 0.0257, 0.0510, 0.0226 compared with the personally identified 0.0009.  Further analysis is needed to better understand the persisting upward skew of women's participation as computer-science authors in 1960–1980 by a minimum of 71% and a maximum of nearly 300%; the 1960–1980 median of 1.78 indicates enduring and substantial 'type two' bias even into this latter period.  It is possible, of course, that other datasets would exhibit a different balance between 'type one' and 'type two' gender bias.

**Big data and "data science"**

The widespread enthusiasm for big data and data science are well known.  Many scholars will be mightily impressed with the Allen Institute's analysis of the 150 million research articles made available in the Semantic Scholar dataset.  Above, I have suggested several ways in which gender bias is created when using gender-predicting software tools for large-scale historical analysis of computer science research.  The striking but anomalous findings from the Allen Institute that women constitute 20 percent of computer-science researchers in the early 1950s certainly did *not* come from the DBLP dataset itself, which had exactly three woman-authored papers in its first four years, and so somewhat fewer than 2 percent women authors.  Could it be the case the larger Semantic Scholar dataset contains a plethora of women computer-science researchers large enough to offset the male-heavy DBLP figures?

A complete answer must await further analysis, but my preliminary assessment is that Semantic Scholar includes wide-ranging and sometimes far-fetched research publications in "computer science," casting some doubt on the entire enterprise.[52]  While DBLP focuses narrowly on the core of computing, the Semantic Scholar opens a far wider aperture.  Some of its "computer science" labeled entries are clearly relevant to computing, others reasonably connect computing with other fields using computers, but rather too many seem the result of some undisciplined imagination or poorly designed scanning and indexing; the guideline that 80 percent of data-science labor is required for preparing clean data to ensure robust results bears repeating.  Here are several far-fetched instances from 1950 that were identified as (solely) computer science: the article "Multiple Primary Carcinoma" reports one dramatic medical case, in *British Medical Journal*, including the medical term 'synchronous' but otherwise it has no computing content;

likewise in the same journal is a piece on the anatomy of the Hounaman monkey. Arnold Sorsby's "Histopathology of the Eye" again in *British Medical Journal* is a book review with nothing on computing. Similarly, Elizabeth F. Gardner's lengthy and searching review of *UCJ: An Orthographic System of Notation and Transcription for Sino-Japanese* appearing in the journal *Language* mentions the linguistics concepts of 'notation', 'special rules', and 'semantic signals' but nothing about computing. And let's not forget the *Indian Medical Gazette's* review of two physiology textbooks.[53] One pointer that we are in treacherous territory is the Semantic Scholar's single entry listing the *title* of one medical-case report (on leeches) while the entry's provided *abstract* instead describes surgically removing glass tumblers from the rectums of a "married man aged 58" and another "married man of 70 years of age."[54] *All* of these wonders were categorized *uniquely* as "computer science."

The phrase "garbage in, garbage out" is typically credited to an Army programmer–instructor in 1957. The corollary in logic is that any argument that proceeds from flawed premises is itself flawed.[55] Here and elsewhere, "dirty data is the most common barrier faced by workers dealing with data."[56]

I have now come full circle with Brian Beaton's "data critique," first mounting an internal investigation of the analytical and statistical methods used in big-data analysis and, above, returning to a more common textual critique based on humanistic approaches and sensibilities. I wish to end on a constructive note, however, since the persisting problem of gender bias in computing needs critique as well as constructive intervention on the part of interdisciplinary researchers in the field.

One such critical and constructive turn recognizes the pernicious gender bias in the immense natural-language corpora used as the basis for machine learning. Computer scientists studying this problem identified the meme "Man is to computer programmer as woman is to homemaker," which received far-reaching publicity.[57] Yet the aims of Bolukbasi et al. were ultimately not critical but constructive: they hope to identify a technical means for correcting the biased word embeddings that led to the sexist-sounding meme. Subsequently, computer science has recognized the adjustment or mitigation of gender, political, and other forms of bias in such word embeddings to be a valid and challenging computer-science problem. Several studies recognize that bias is typically not one-dimensional (say, politics or gender alone) but often multi-dimensional.[58] For humanists exploring the intersectionality of bias across race, class, and gender, we have a more receptive audience in computer science than we might usually imagine.[59]

In this regard, gender bias in computing is an urgent interdisciplinary field that needs attention by historians, humanities scholars, and well-informed technical colleagues.[60] This article shows that combining traditional humanistic critique with informed inspection of big data methods is feasible, productive, and fruitful. Some big data researchers (as noted above) need to be reminded that their studies require serious attention to creating robust and meaningful sources of data; sometimes the challenging 80/20 split—preparation of data compared with analysis of data—is wrongly identified as an unfortunate problem to be somehow overcome for data science to flourish.  I believe this is misguided, as the errant data lodged in Semantic Scholar indicates.  Robust insight into gender bias and other serious socio-technical problems facing our society is an urgent problem requiring deeper integration of humanistic, social science, and computer science researchers.  Come to think of it: is there any pressing society problem—such as climate change, structural racism, or economic challenges—that isn't exactly the same?

This is a pre-copyedited version of an article accepted for publication in Information & Culture 57 no. 3 (2022): 283-306 following peer review. The definitive publisher-authenticated version is available from University of Texas PressENDNOTES

[1] See, for instance, Carol Frieze, "Diversifying the Images of Computer Science: Undergraduate Women take on the Challenge," *SIGCSE Bulletin* 37 no. 1 (2005): 397-400 at doi.org/10.1145/1047124.1047476; Sapna Cheryan, Victoria C. Plaut, Caitlin Handron, and Lauren Hudson, "The Stereotypical Computer Scientist: Gendered Media Representations as a Barrier to Inclusion for Women," *Sex Roles: A Journal of Research* 69 nos. 1-2 (2013): 58-71 at doi.org/10.1007/s11199-013-0296-x; Matthew Kay, Cynthia Matuszek, and Sean A. Munson, "Unequal Representation and Gender Stereotypes in Image Search Results for Occupations," In Proceedings of the 33rd Annual ACM Conference on Human Factors in Computing Systems (CHI '15) Association for Computing Machinery (2015): 3819-28 at doi.org/10.1145/2702123.2702520; and Jenna Cryan, Shiliang Tang, Xinyi Zhang, Miriam Metzger, Haitao Zheng, and Ben Y. Zhao, "Detecting Gender Stereotypes: Lexicon vs. Supervised Learning Methods," In Proceedings of the 2020 CHI Conference on Human Factors in Computing Systems (CHI '20) Association for Computing Machinery (2020): 1-11 at doi.org/10.1145/3313831.3376488

[2] Cited in the "guidelines and warnings" for Lincoln Mullen, "gender: Predict Gender from Names Using Historical Data," R package version 0.5.4.1000 (2021) at github.com/lmullen/gender (accessed October 2021).

[3] Cameron Blevins and Lincoln Mullen, "Jane, John ... Leslie? A Historical Method for Algorithmic Gender Prediction," *Digital Humanities Quarterly* 9 no. 3 (2015): n.p. (quotes paragraph 5) at www.digitalhumanities.org/dhq/vol/9/3/000223/000223.html

[4] See Steve Lohr, "For Big-Data Scientists, 'Janitor Work' Is Key Hurdle to Insights," *New York Times* (August 17, 2014) at www.nytimes.com/2014/08/18/technology/for-big-data-scientists-hurdle-to-insights-is-janitor-work.html; Armand Ruiz, "The 80/20 Data Science Dilemma," *InfoWorld* (September 26, 2017) at www.infoworld.com/article/3228245/the-80-20-data-science-dilemma.html; and Ihab F. Ilyas and Xu Chu, *Data Cleaning* (New York: Association for Computing Machinery, 2019) at doi.org/10.1145/3310205.

[5] Janet Abbate, *Recoding Gender: Women's Changing Participation in Computing* (Cambridge: MIT Press, 2012), quotes p. 3.

[6] Penelope Eckert, "Coding for Gender and Sexuality," NSF–Linguistic Data Consortium workshop on Coding Sociolinguistic Corpora (Linguistic Society of America Annual Meeting, Portland, 2012) at www.ldc.upenn.edu/sites/www.ldc.upenn.edu/files/eckert-paper.pdf

[7] A sample of these productive scholars' publications sustained over decades includes Margaret Rossiter's three volumes on *Women Scientists in America* (Baltimore: Johns Hopkins University Press, 1982, 1995, 2012); Mary Frank Fox, Deborah G. Johnson, and Sue V. Rosser, eds. *Women, Gender, and Technology* (Urbana: University of Illinois Press, 2006); Sally Gregory Kohlstedt and Helen Longino, eds., "Women, Gender, and Science: New Directions," *Osiris* Vol. 12 (1997); and Ruth Oldenziel, *Making Technology Masculine: Men, Women, and Modern Machines in America, 1870-1945* (Amsterdam: Amsterdam University Press, 1999).

[8] Margaret W. Rossiter, *Women Scientists in America: Forging a New World since 1972* (Baltimore: Johns Hopkins University Press, 2012), quote p. 41.

[9] MIT Press (2006), quote p. ix.

[10] Computing programs at Harvey Mudd, Carnegie Mellon, and University of California–Berkeley are seen as success stories according to Sarah McBride, "Glimmers of Hope for Women in the Male-Dominated Tech Industry," Bloomberg Technology (8 March 2018) at www.bloomberg.com/news/articles/2018-03-08/glimmers-of-hope-for-women-in-the-male-dominated-tech-industry

[11] Cathy O'Neil, *Weapons of Math Destruction: How Big Data Increases Inequality and Threatens Democracy* (New York: Crown, 2016); Meredith Broussard, *Artificial Unintelligence: How Computers Misunderstand the World* (Cambridge: MIT Press, 2018); Ruha Benjamin, *Race After Technology: Abolitionist Tools for the New Jim Code* (2019); Safiya Umoja Noble, *Algorithms of Oppression: How Search Engines Reinforce Racism* (New York: New York University Press, 2018). An overview is Megan Garcia, "Racist in the Machine: The Disturbing Implications of Algorithmic Bias," *World Policy Journal* 33 no. 4 (Winter 2016/2017): 111-117

[12] Christine T. Wolf, review of *Invisible Women* in *Information & Culture* 54 no. 3 (2019): 400-402

[13] Thomas S. Mullaney, Benjamin Peters, Mar Hicks, and Kavita Philip, eds., *Your Computer Is on Fire* (Cambridge: MIT Press, 2021)

[14] Citations in Thomas J. Misa, "Gender Bias in Computing," in William Aspray, ed., *Historical Studies in Computing, Information, and Society* (Springer, 2019), 113-133 on 120-122 (notes 22-33) at doi.org/10.1007/978-3-030-18955-6_6

[15] Core citations are W. Barkley Fritz, "The Women of ENIAC," *IEEE Annals of the History of Computing* 18 no. 3 (1996): 13-28 at doi.org/10.1109/85.511940; Jennifer S. Light, "When Computers Were Women," *Technology and Culture* 40 no. 3 (1999): 455-483, quotes on 455 and 469 at www.jstor.org/stable/25147356; Nathan Ensmenger, "Making Programming Masculine," in Thomas J. Misa, ed., *Gender Codes: Why Women Are Leaving Computing* (John Wiley, 2010), 115-141 at doi.org/10.1002/9780470619926.ch6 ; and Nathan Ensmenger, "'Beards, Sandals, and Other Signs of Rugged Individualism': Masculine Culture within the Computing Professions," *Osiris* 30 (2015): 38-65 at doi.org/10.1086/682955. For his claim of 50 percent women programmers, see Ensmenger at https://web.archive.org/web/20180105182302/homes.soic.indiana.edu/nensmeng/files/ensmenger-gender.pdf on p. 2.

[16] José María Cavero, Belén Vela, Paloma Cáceres, Carlos Cuesta, and Almudena Sierra-Alonso, "The Evolution of Female Authorship in Computing Research," *Scientometrics* 103 (2015): 85-100, on p. 89, at DOI 10.1007/s11192-014-1520-3; J. McGrath Cohoon, Sergey Nigai, and Joseph 'Jofish' Kaye, "Gender and Computing Conference Papers," *Communications of the ACM* 54 no. 8 (2011): 72-80 at doi.org/10.1145/1978542.1978561; Thomas J. Misa, "Dynamics of Gender Bias in Computing," *Communications of the ACM* 64 no. 6 (June 2021): 76-83 at doi.org/10.1145/3417517

[17] Abbate, *Recoding Gender*, quote p. 1; Thomas Haigh and Paul E. Ceruzzi, *A New History of Modern Computing* (Cambridge: MIT Press, 2021), quote p. 63.

[18] Christianne Corbett and Catherine Hill, *Solving the Equation: The Variables for Women's Success in Engineering and Computing* (Washington DC: AAUW, 2015), p. 9 figure 1.

[19] The elite Society of Actuaries remained a male bastion well into the 1960s, but it is unlikely to reflect the much larger actuarial workforce. See Jillian Emberg, "A Study of Women Working in the Actuarial Field," (BA Thesis, Bryant University, 2012) at web.archive.org/web/20210507185232/https://digitalcommons.bryant.edu/cgi/viewcontent.cgi?article=1008&context=honors_mathematics

[20] Lucy Lu Wang, Gabriel Stanovsky, Luca Weihs, and Oren Etzioni, "Gender Trends in Computer Science Authorship," *Communications of the ACM* 64 no. 3 (2021): 78-84, figure 4 on page 83, at doi.org/10.1145/3430803

[21] For a population of around a dozen, see Trudi Bellardo Hahn, Diane L. Barlow, "Women Pioneers in the Information Sciences," *Libraries & the Cultural Record* 45 no. 2 (2010): 163-166. Compare the analysis of 57 black computing professionals in R. Arvid Nelsen, "Race and Computing: The Problem of Sources, the Potential of Prosopography, and the Lesson of *Ebony* Magazine," *IEEE Annals of the History of Computing* 39 no. 1 (2017): 29-51 at doi: 10.1109/MAHC.2016.11

22 Laine Nooney, Kevin Driscoll, Kera Allen, "From Programming to Products: *Softalk* Magazine and the Rise of the Personal Computer User," *Information & Culture* 55 no. 2 (2020): pp. 105-129, quotes pp. 109, 111, 125, and 129 note 52.

23 For assessment of gender-prediction tools: Lucía Santamaría and Helena Mihaljević, "Comparison and Benchmark of Name-to-Gender Inference Services," *PeerJ Computer Science* (July 16, 2008), quote p. 2 (best performer), at dx.doi.org/10.7717/peerj-cs.156; Fariba Karimi, Claudia Wagner, Florian Lemmerich, Mohsen Jadidi, and Markus Strohmaier, "Inferring Gender from Names on the Web: A Comparative Evaluation of Gender Detection Methods," In Proceedings of the 25th International Conference Companion on World Wide Web (WWW '16 Companion). International World Wide Web Conferences Steering Committee, Republic and Canton of Geneva, CHE, 5-54 at doi.org/10.1145/2872518.2889385 ; Hua Zhao and Fairouz Kamareddine, "Advance Gender Prediction Tool of First Names and its Use in Analysing Gender Disparity in Computer Science in the UK, Malaysia and China," 2017 International Conference on Computational Science and Computational Intelligence (CSCI), Las Vegas, NV, USA, 2017, pp. 222-227, at doi.org/10.1109/CSCI.2017.35 ; Foad Hamidi, Morgan Klaus Scheuerman, and Stacy M. Branham, "Gender Recognition or Gender Reductionism? The Social Implications of Embedded Gender Recognition Systems," In Proceedings of the 2018 CHI Conference on Human Factors in Computing Systems (CHI '18). Association for Computing Machinery, New York, NY, USA, Paper 8, 1-13 at doi.org/10.1145/3173574.3173582 ; Stefan Krüger and Ben Hermann, "Can an Online Service Predict Gender? On the State-of-the-Art in Gender Identification From Texts," In Proceedings of the 2nd International Workshop on Gender Equality in Software Engineering (GE '19). IEEE Press, 13-16 at doi.org/10.1109/GE.2019.00012

24 Blevins and Mullen (2015), quote paragraph 14.  Their R software tool "gender: Predict Gender from Names Using Historical Data," credited to Lincoln Mullen, Cameron Blevins, and Ben Schmidt, is located at rdrr.io/cran/gender/ and github.com/lmullen/gender (October 2021).

25 A manuscript reviewer suggested this article, in addition to Gender-API, NamSor, and Genderizer.io, also compare the gender prediction tool proposed in Blevins and Mullen (2015).  Their R-based software package can utilize three or more different historical datasets, and its results depend on which ones are chosen; so a direct comparison is not simple to arrange.

26 See Muneera Bano and Didar Zowghi, "Gender disparity in the governance of software engineering conferences," In Proceedings of the 2nd International Workshop on Gender Equality in Software Engineering (GE '19) IEEE Press, 21-24 at https://doi.org/10.1109/GE.2019.00016 ; Abdulhakim Qahtan, Nan Tang, Mourad Ouzzani, Yang Cao, and Michael Stonebraker, "Pattern functional dependencies for data cleaning," *Proceedings of the VLDB Endowment* 13 no. 5 (January 2020): 684-697 at https://doi.org/10.14778/3377369.3377377;  Sandra Mattauch, Katja Lohmann, Frank Hannig, Daniel Lohmann, and Jürgen Teich, "A Bibliometric Approach for Detecting the Gender Gap in Computer Science," *Communications of the ACM* 63 no. 5 (2020): 74-80, quotes 76, 78, at doi.org/10.1145/3376901

27 An early computer-science gender assessment also manually verified the output of its prototype gender-prediction software tool (Genderyzer, "an experimental research tool for determining the gender mix of a list of names"); see J. McGrath Cohoon, Sergey Nigai, and Joseph 'Jofish' Kaye, "Gender and Computing Conference Papers," *Communications of the ACM 54* no. 8 (2011): 72-80. Genderyzer is currently located at jofish.com/cgi-bin/genderyze.py but no longer actively supported according to a note in the Python script dated December 2006.

28 Sandra Mattauch, Katja Lohmann, Frank Hannig, Daniel Lohmann, and Jürgen Teich, "A Bibliometric Approach for Detecting the Gender Gap in Computer Science," *Communications of the ACM* 63 no. 5 (2020): 74-80, quotes 76, 78, at doi.org/10.1145/3376901

This is a pre-copyedited version of an article accepted for publication in Information & Culture 57 no. 3 (2022): 283-306 following peer review. The definitive publisher-authenticated version is available from University of Texas Press[29] Wang et al., "Gender Trends in Computer Science Authorship," quote 78.

[30] Gender API labels as "accuracy" simply the number of female (or male) instances of the name divided by the total (female and male) instances; it works well, on a frequentist interpretation of probability, where there are large numbers: its "accuracy" is 98% for both 'John' (139,211 samples) and 'Anna' (184,558 samples), likely reasonable figures. But when samples become small enough, Gender API's computed "accuracy" can, nonsensically, go up. So 'Atwell' has a 91% "accuracy" of being male, with 11 samples (so one back-calculates 10 male and 1 female instances). Both 'Christian' and 'Giampio' (respectively 48,300 and just 4 samples) are each computed to be 100% "accurate" as male names. This is an error in statistical sampling, since where n = 4 you cannot properly have 100% "accuracy" in the sample. Other names Gender API computes to have 100% "accuracy" with tiny samples of 2 to 4 include Kumpati, Tachen, Yow-Yieh, Shaler, Belur, and Acheson.

[31] The Python-based "Gender-Guesser" API labels non-binary names including "'unknown' (name not found), 'andy' (androgynous), 'male', 'female', 'mostly_male', or 'mostly_female'." See pypi.org/project/gender-guesser/

[32] Wang, et al., "Gender Trends," quotes 78, 79

[33] Brian Beaton, "How to Respond to Data Science: Early Data Criticism by Lionel Trilling," *Information & Culture* 51 no. 3 (2016): 352-372, quote 353, at doi.org/10.1353/lac.2016.0014. In a supportive vein, William Aspray acknowledges inspiration from "methods from critical theory and cultural studies" in his article "Information History: Searching for Identity," *Information & Culture* 54 no. 1 (2019): 69-75 quote 74.

[34] Corina Koolen and Andreas van Cranenburgh, "These are not the Stereotypes You are Looking For: Bias and Fairness in Authorial Gender Attribution," *Proceedings of the First Workshop on Ethics in Natural Language Processing* (Valencia, Spain, April 4, 2017): 12-22, quote 14, at DOI 10.18653/v1/W17-1602

[35] Ann K. Boulis and Jerry A. Jacobs, *The Changing Face of Medicine: Women Doctors and the Evolution of Health Care in America Book* (Cornell University Press, 2008), page 20, figure 2.2 "New MD Degrees by Gender, 1950–2004" citing NCES's Digest of Education Statistics.

[36] Amy Sue Bix, "From 'Engineeresses' to 'Girl Engineers' to 'Good Engineers': A History of Women's U.S. Engineering Education," *National Women's Studies Association Journal* 16 no. 1 (2004): 27-49, quote 27, at muse.jhu.edu/article/168384.

[37] My manual method using year-specific SSA data is thus similar to Blevins and Mullen's R software tool using the SSA dataset, noted above. Laplace once hailed logarithms "as doubling the life of the astronomer" as quoted in Jack Oliver, "The Birth of Logarithms," *Mathematics in School* 29 no. 5 (2000): 9-13, on p. 12, at www.jstor.org/stable/30215439. Similarly, I acknowledge my son's grep script speeding my look ups of SSA data.

[38] Thomas J. Misa, "Gender Bias in Computing," in William Aspray, ed., *Historical Studies in Computing, Information, and Society* (Springer, 2019), 113-133 at doi.org/10.1007/978-3-030-18955-6_6; and idem, "Dynamics of Gender Bias in Computing," *Communications of the ACM* 64 no. 6 (June 2021): 76-83 at doi.org/10.1145/3417517

[39] DBLP's very first years, beginning with 1936 (the year of Alonzo Church's celebrated "Note on the Entscheidungsproblem") are heavily dominated by *Journal of Symbolic Logic* with sporadic entries from *Annales des Télécommunications* (1946) and *Bell System Technical Journal* (1948-49).

[40] Purposely hiding one's gender can be a deliberate step for some women PhD students in engineering, according to Shelley K. Erickson, "Women Ph.D. Students in Engineering and a Nuanced Terrain: Avoiding and Revealing Gender," *Review of Higher Education* 35 no. 3 (2012): 355-374 at doi.org/10.1353/rhe.2012.0019

41 Kamil Wais, "Gender Prediction Methods Based on First Names with genderizeR," *The R Journal* 8 no. 1 (August 2016) at journal.r-project.org/archive/2016-1/wais.pdf; Jevin D. West, Jennifer Jacquet, Molly M. King, Shelley J. Correll, and Carl T. Bergstrom, "The Role of Gender in Scholarly Authorship," PLOS ONE 8 no. 7 (July 22, 2013) at doi.org/10.1371/journal.pone.0066212.

42 Suppression of edge cases seems endemic. In an otherwise careful comparative machine-learning DH experiment, Koolen and van Cranenburgh (cited above) oddly decided to "leave out authors of unknown or multiple genders" (quote p. 18). Another machine-learning DH study, in order to "reduce the feature set and transform feature values into entities a prediction model can work with" took steps to "remove all the geo-references," "merge all the Session Start and Session End columns into one [sic] POSIX date column," as well as "factorized (value-indexed) the 9 categorical features" on a path to "reduce the overall number of significant features to 5" for machine-learning analysis; see Tobias Blanke, "Predicting the Past," *Digital Humanities Quarterly* 12 no. 2 (2018), quotes on paragraphs 26-27 at digitalhumanities.org:8081/dhq/vol/12/2/000377/000377.html . Unaccountably, the Allen Institute's Wang et al. (cited above) admit: "We also filter out first names that occur less than 10 times in our overall corpus, to reduce the number of API calls to manageable numbers." Remarkably, one author simply "removed from his corpus the 'troublemakers' who disturbed his results too deeply because it is 'notoriously too difficult to gender classify'," according to Laura Mandell, "Gender and Cultural Analytics: Finding or Making Stereotypes?" in Matthew K. Gold and Lauren F. Klein, eds. *Debates in the Digital Humanities* (University of Minnesota Press, 2019), 1-22, quote p. 8. Explaining a similar discard, "we obtained better results when using only [gender] unambiguous names," (and disregarding gender ambiguous ones) state José María Cavero, Belén Vela, Paloma Cáceres, Carlos Cuesta, and Almudena Sierra-Alonso, "The Evolution of Female Authorship in Computing Research," *Scientometrics* 103 (2015): 85-100, quote p. 88, at DOI 10.1007/s11192-014-1520-3

43 Eduardo Graells-Garrido, Mounia Lalmas, and Filippo Menczer, "First Women, Second Sex: Gender Bias in Wikipedia," Proceedings of the 26th ACM Conference on Hypertext & Social Media (HT '15), Association for Computing Machinery, New York, NY, USA, 2015, 165-174 at doi.org/10.1145/2700171.2791036

44 For instance, computer science professor A. J. Kfoury [Assaf J] has a personal website [www.bu.edu/prsocial/profile/assaf-j-kfoury/](www.bu.edu/prsocial/profile/assaf-j-kfoury/) using male pronouns; IBM researcher A. K. Gaind [Arun] has an author biography including 'Mr. Gaind' and male pronouns; B. L. Houseman [Barton], a 'professor emeritus' in chemistry was interviewed, with photo and use of male pronouns, for a Goucher College alumni magazine; and 'professor emeritus' in computer science C. K. Wong [Chak-Kuen] has a personal website at appsrv.cse.cuhk.edu.hk/~wongck with photo and use of male pronouns.

45 Mattauch et al (op cit.) explicitly discuss their "removal of Asian names" while acknowledging this as a "potential source of error of our approach," on pp. 77 and 80; while Wang et al. note "East Asian first names, when romanized, are more gender ambiguous" on p. 82.

46 One analysis of DBLP authors 1936–2010 found women to vary year-by-year between 0 and 5.5 percent of total authors during 1942–54, with (my estimate from their graph) an average of approximately 1.8 percent during 1949-54; see Cavero et al. "Evolution of Female Authorship in Computing Research," op cit.

47 The basic statistics sampling equation, akin to physicists' $F = MA$, prominently appears in Mattauch et al.'s 2020 *Communications of the ACM* article on page 78. For a useful textbook, see Sharon L. Lohr, *Sampling: Design and Analysis* (Pacific Grove, CA : Duxbury Press, 1999), where this equation is derived on page 40.

This is a pre-copyedited version of an article accepted for publication in Information & Culture 57 no. 3 (2022): 283-306 following peer review.  The definitive publisher-authenticated version is available from University of Texas Press48 Authors per paper averaged 1.4 during 1955-64, rising to a range 2.5–4.2 (by sub-field) during 2005-14, according to João Fernandes and Miguel Monteiro, "Evolution in the Number of Authors of Computer Science Publications," *Scientometrics* 110 (2016): 1-11 at DOI: 10.1007/s11192-016-2214-9.

49 For example, Grems obituary-memoir at schenectadyhistorical.org/admin/wp-content/uploads/2013/08/Nov-Dec-1998.pdf; "Florence MacWilliams, A Mathematician, 73," *New York Times* (May 31, 1990) at www.nytimes.com/1990/05/31/obituaries/florence-macwilliams-a-mathematician-73.html; amp.en.google-info.in/58707434/1/joan-francioni.html

50 "In general, names tend to shift from being predominantly male to being predominantly female; the shift rarely happens in the other direction," state Blevins and Mullen in note 2 (cited above).

51 For the computation of gender "ratios," I exactly "matched" all individuals by setting aside (a) the personally identified "initials only" individuals, since they didn't have first names for the gender-predicting software tools to analyze; and (b) the individuals whose first names were identified by the software tools but who could not be personally identified and were not in the SSA database of names.  The figures for overall women's participation rates included this partially matched data.

52 My preliminary assessment was done on 9 May 2021; Semantic Scholar periodically issues new builds of the dataset.

53 See, respectively, doi.org/10.1136/bmj.2.4676.462; doi.org/10.1136/BMJ.2.4670.90-B; www.bmj.com/content/2/4673/254.2; doi.org/10.2307/410083; doi.org/10.1136/bmj.2.5041.396-c

54 The *BMJ* page was sloppily scanned and poorly analyzed: see the *source page* doi.org/10.1136/BMJ.2.4684.891-B *resulting in* the split title–abstract entry at www.semanticscholar.org/paper/Haematemesis-from-Leeches-Cunningham/d6af39735e2384fc8d531fb4d6b0c54595949e13

55 "Work with New Electronic 'Brains' Opens Field For Army Math Experts," *Hammond [Indiana] Times* (10 November 1957) at web.archive.org/web/20210309040239/https://www.newspapers.com/clip/50687334/the-times/

56 Ilyas and Chu, *Data Cleaning* (cited above) quote p. 1.

57 Tolga Bolukbasi, Kai-Wei Chang, James Zou, Venkatesh Saligrama, and Adam Kalai, "Man is to Computer Programmer as Woman Is To Homemaker? Debiasing Word Embeddings," (2016) available at arxiv.org/abs/1607.06520v1 and elsewhere.

58 Marzieh Babaeianjelodar, Stephen Lorenz, Josh Gordon, Jeanna Matthews, and Evan Freitag, "Quantifying Gender Bias in Different Corpora," In Companion Proceedings of the Web Conference 2020 (WWW '20) Association for Computing Machinery, New York, NY, 2020, 752-759 at doi.org/10.1145/3366424.3383559; Joshua Gordon, Marzieh Babaeianjelodar, and Jeanna Matthews, "Studying Political Bias via Word Embeddings," Companion Proceedings of the Web Conference 2020 (WWW '20) Association for Computing Machinery, New York, NY, 2020, 760-764 at doi.org/10.1145/3366424.3383560; Ninareh Mehrabi, Thamme Gowda, Fred Morstatter, Nanyun Peng, and Aram Galstyan, "Man is to Person as Woman is to Location: Measuring Gender Bias in Named Entity Recognition," In Proceedings of the 31st ACM Conference on Hypertext and Social Media (HT '20) Association for Computing Machinery, New York, NY, 2020, at doi.org/10.1145/3372923.3404804. Mitigation measures are proposed and evaluated in Mascha Kurpicz-Briki and Tomaso Leoni, "A World Full of Stereotypes? Further Investigation on Origin and Gender Bias in Multi-Lingual Word Embeddings," *Frontiers in Big Data* (3 June 2021) 4:625290. doi: 10.3389/fdata.2021.625290 ; and Helena Mihaljevic, Marco Tullney, Lucía Santamaría, and Christian Steinfeldt, "Reflections on Gender Analyses of Bibliographic Corpora," *Frontiers in Big Data* (28 August 2019) 2:29. doi: 10.3389/fdata.2019.00029

[59] See (e.g.) Joy Buolamwini and Timnit Gebru, "Gender Shades: Intersectional Accuracy Disparities in Commercial Gender Classification," Proceedings of Machine Learning Research 81 (2018): 1-15 at proceedings.mlr.press/v81/buolamwini18a/buolamwini18a.pdf; and Michelle Trim, "Essentialism is the Enemy of the Good: How the Myth of Objectivity is Holding Computing Back," *SIGCAS Computers and Society* 49 no. 2 (2020): 11-13 at doi.org /10.1145/3447903.3447908; Roopika Risam, "Beyond the Margins: Intersectionality and the Digital Humanities," *Digital Humanities Quarterly* 9 no. 2 (2015) at digitalhumanities.org:8081/dhq/vol/9/2/000208/000208.html.  Buolamwini and Gebru (then at Microsoft, later at Google, and a *cause célèbre* for being fired by Google in December 2020) cite Cathy O'Neil 2017 [see above], while Trim might have cited historian Amy E. Slaton's *Race, Rigor, and Selectivity in U.S. Engineering: The History of an Occupational Color Line* (Cambridge: Harvard University Press, 2010).

[60] For enthusiastic forecasts for digital humanities, see Thomas Bartscherer and Roderick Coover, eds., *Switching Codes: Thinking Through Digital Technology in the Humanities and the Arts* (Chicago: University of Chicago Press, 2011), which conjures up "technologies that are so radically transforming [scholars'] fields" (quote p. 2); William J. Turkel, Shezan Muhammedi, and Mary Beth Start, "Grounding Digital History in the History of Computing," *IEEE Annals of the History of Computing* 36, no. 2 (2014): 72-75 at doi: 10.1109/MAHC.2014.21. Contrast the insightful contribution of Marc Weber, "Self-Fulfilling History: How Narrative Shapes Preservation of the Online World," *Information & Culture* 51 no. 1 (2016): 54-80 at doi:10.1353/lac.2016.0003.